\documentclass[conference]{IEEEtran}

\usepackage[utf8]{inputenc}
\usepackage[T1]{fontenc}

\usepackage{booktabs}
\usepackage{tabularx}
\usepackage{paralist}

\usepackage{flushend}

\usepackage{cite}
\usepackage{amsmath,amssymb,amsfonts}
\usepackage{algorithmic}
\usepackage{graphicx}
\usepackage{textcomp}
\usepackage{xcolor}
\usepackage{tikz}

\usepackage{framed}

\usepackage{tcolorbox}
\tcbuselibrary{theorems}
\usepackage{fontawesome}
\usepackage{url}

\newtcbtheorem{system}{\faGears{} System Prompt}
{sharp corners=all,colback=gray!5,colframe=blue!35!black,fonttitle=\bfseries,left=0.1cm,right=0.1cm,top=0.1cm,bottom=0.1cm}{th}

\newcommand\copyrighttext{%
  \footnotesize \textcopyright~2025 IEEE. Personal use of this material is permitted. Permission from IEEE must be obtained for all other uses, in any current or future media, including reprinting/republishing this material for advertising or promotional purposes, creating new collective works, for resale or redistribution to servers or lists, or reuse of any copyrighted component of this work in other works.}
\newcommand\copyrightnotice{%
\begin{tikzpicture}[remember picture,overlay]
\node[anchor=south,yshift=10pt] at (current page.south) {\fbox{\parbox{\dimexpr\textwidth-\fboxsep-\fboxrule\relax}{\copyrighttext}}};
\end{tikzpicture}%
}

\begin{document}

\title{LLMREI: Automating Requirements Elicitation Interviews with LLMs}


\author{
\IEEEauthorblockN{Alexander Korn\IEEEauthorrefmark{1}, Samuel Gorsch\IEEEauthorrefmark{2}, Andreas Vogelsang\IEEEauthorrefmark{1}}
\IEEEauthorblockA{\IEEEauthorrefmark{1}paluno -- The Ruhr Institute for Software Technology \\ University of Duisburg-Essen \\ Essen, Germany\\
Email: \{alexander.korn, andreas.vogelsang\}@uni-due.de}
\IEEEauthorblockA{\IEEEauthorrefmark{2}University of Cologne \\ Cologne, Germany\\
Email: samuel@gorsch.de}
}

\maketitle

\begin{abstract}
Requirements elicitation interviews are crucial for gathering system requirements but heavily depend on skilled analysts, making them resource-intensive, susceptible to human biases, and prone to miscommunication. Recent advancements in Large Language Models present new opportunities for automating parts of this process. This study introduces LLMREI, a chat bot designed to conduct requirements elicitation interviews with minimal human intervention, aiming to reduce common interviewer errors and improve the scalability of requirements elicitation. We explored two main approaches, zero-shot prompting and least-to-most prompting, to optimize LLMREI for requirements elicitation and evaluated its performance in 33 simulated stakeholder interviews. A third approach, fine-tuning, was initially considered but abandoned due to poor performance in preliminary trials. Our study assesses the chat bot's effectiveness in three key areas: minimizing common interview errors, extracting relevant requirements, and adapting its questioning based on interview context and user responses. Our findings indicate that LLMREI makes a similar number of errors compared to human interviewers, is capable of extracting a large portion of requirements, and demonstrates a notable ability to generate highly context-dependent questions. We envision the greatest benefit of LLMREI in automating interviews with a large number of stakeholders.
\end{abstract}


\copyrightnotice
\section{Introduction}

Requirements Engineering (RE) is a critical discipline within software development, ensuring to align software products with stakeholder needs by involving gathering, analyzing, and reviewing specifications~\cite{chakraborty_role_2012}.

The elicitation phase is fundamental to Requirements Engineering, as it lays the groundwork for accurately capturing stakeholder needs. This phase frequently relies on stakeholder interviews, which are widely regarded as an effective technique for gathering requirements. However, conducting interviews is highly resource-intensive~\cite{Zowghi}, requiring significant time, coordination, and the involvement of skilled analysts to guide discussions and interpret information accurately~\cite{pacheco_requirements_2018}. In addition to the high resource demands, interviews are susceptible to human bias, and misinterpretation, which can result in incomplete or unclear requirements~\cite{pacheco_requirements_2018, ferrari_ambiguity_2016, zalewski_cognitive_2020}. The considerable effort required, combined with these inherent limitations, makes requirements elicitation an attractive candidate for investigating opportunities for LLM-based support and automation.

Recent advances in Large Language Models (LLMs) introduce the potential for augmenting RE processes, especially in requirements elicitation~\cite{delima_automatic_2021,liu_artificial_2022}. Models such as GPT have demonstrated substantial capabilities in understanding and generating language~\cite{zhao2024surveylargelanguagemodels}. In requirements elicitation, LLMs could assist by automating interviews~\cite{Vogelsang2025-LLM, delima_automatic_2021} and reducing human-related challenges, such as bias and miscommunication. However, the effectiveness of LLMs in the conversational aspects of RE remains under-researched, as requirements gathering requires nuanced human interaction to interpret stakeholder intentions accurately~\cite{arora_advancing_2023, haq_use_2019, Vogelsang2025-LLM, ling_domain_2024}.

LLMs offer possibilities to streamline RE by enhancing interviewer consistency and scalability. However, domain-specific customization is necessary for LLMs to be effective in RE. In this paper, we explore the potential of an LLM-based chat bot, \textit{LLMREI}, to conduct RE interviews autonomously, addressing typical challenges faced by human interviewers. As the LLMREI chat bot is designed to enhance the accuracy and scalability of requirements interviews, this study examines its effectiveness in reducing interviewer errors, dynamically adapting the flow of questions based on stakeholder responses, and eliciting a greater number of relevant requirements. By customizing LLMs through prompt engineering and fine-tuning, we evaluate different approaches to optimize LLMREI for requirements interviewing. We primarily focus on using pre-existing LLMs, GPT-4o specifically, to compare performance.

While LLMREI has potential applications in RE training, this study focuses solely on interview automation. Our results show that LLMREI makes a similar number of common mistakes during interviews as a human interviewer, while being able to elicit up to 73.7~\% of all requirements, proving the capability of LLMs to conduct requirements interviews.

We are convinced that human-led interviews are still valuable and necessary to cope with highly sensitive or complex elicitation scenarios,
where interpersonal dynamics play a crucial role in uncovering
unspoken requirements. However, we envision benefits from using LLMREI in the early phases of requirements elicitation. By enabling (semi-)automated interviews with a large number of stakeholders, we see the potential to efficiently gather a substantial portion of initial requirements with minimal human effort. These preliminary requirements can then be used to provide human analysts with a clearer understanding of the project context and stakeholder needs, allowing subsequent manual interviews to be more focused and effective in uncovering detailed, critical, or nuanced requirements that automated methods may overlook.

Our paper makes the following contributions:
\begin{itemize}
    \item We propose and test three approaches to \textbf{LLM-based requirements elicitation interview bots}. Two use prompt engineering techniques: a zero-shot version that uses GPT-4o, and a least-to-most prompting version that uses GPT-4o with extensive interview guidelines reused from existing literature. The third approach utilizes fine-tuning a GPT-3.5 model with transcripts of real interviews.
    \item We conduct an \textbf{evaluation of the prompt engineering approaches} based on a set of 33 simulated stakeholder interviews. We assess the performance of the two variants by manually verifying whether the bots make mistakes common for elicitation interviews. Furthermore, we evaluate the number of relevant requirements elicited and how much the LLMs adapt their questions to the conversational context and user responses.
    \item We provide a comprehensive \textbf{replication package} that contains the study instruments, the raw interview data, our ground truth labels, and the code we used to analyze the data.\footnote{\url{https://doi.org/10.5281/zenodo.14988928}}
\end{itemize}

This research contributes to understanding how LLMs can assist or even automate aspects of the requirements elicitation process, potentially complementing traditional human-led RE activities.

\section{Background and Related Work}
\subsection{Requirements Elicitation Interviews}

Interviews are a widely used technique in RE for eliciting stakeholder needs and preferences directly, offering a detailed understanding of user requirements~\cite{ferrari_ambiguity_2016}. As described by Christel and Kang~\cite{christel_issues_1992}, the interview process generally involves four main steps: 

\begin{enumerate}
\item \textbf{Preparation:} Identifying stakeholders, formulating questions, and defining interview goals. Proper preparation is critical to ensure the interview yields relevant, goal-aligned information.

\item \textbf{Conducting Interviews:} Analysts gather insights from stakeholders either through structured (predefined questions) or unstructured (open-ended) formats. The structure chosen influences the depth and quality of the information obtained.

\item \textbf{Documentation:} Documenting findings accurately is essential to avoid misinterpretation or data loss. Without standardized documentation, inconsistencies in data are more likely.

\item \textbf{Analysis and Integration:} The final step requires synthesizing and reconciling diverse stakeholder views into a unified set of requirements.

\end{enumerate}

Interview structures range from fully structured to completely unstructured. Structured interviews follow a strict set of questions, offering consistency but limited flexibility. Unstructured interviews, in contrast, are open-ended and provide broader insights but are time-consuming and less consistent. Semi-structured interviews offer a balance, combining guided questions with flexibility, making them particularly useful for obtaining both high-level and detailed requirements~\cite{pacheco_requirements_2018}.

While interviews can yield rich qualitative data and enable exploration of unforeseen topics, they also pose several challenges:

\begin{itemize}    
    \item \textbf{Inconsistencies in Interpretation:} Stakeholder interpretations vary due to unique perspectives and terminology, complicating the synthesis of requirements into a unified set. This can result in misaligned expectations~\cite{sharma_requirements_2014}.
    
    \item \textbf{Cognitive Bias:} Analysts may unconsciously favor information that aligns with their pre-existing knowledge, potentially leading to assumptions about user needs rather than capturing an accurate articulation of stakeholder requirements~\cite{hadar_role_2014}.
    
    \item \textbf{Ambiguity in Communication:} Ferrari et al.~\cite{ferrari_ambiguity_2016} identify four types of ambiguity (general ambiguity, multiple understandings, incorrect disambiguation, and correct disambiguation). While ambiguity can sometimes help uncover tacit knowledge, it often causes misinterpretation, impacting the clarity and accuracy of requirements.
    
    \item \textbf{Resource Intensiveness:} Interviews demand significant time and expertise from analysts, limiting scalability and introducing variability in quality. The high skill level required further reduces scalability, which can lead to the exclusion of critical perspectives if the interview volume is constrained~\cite{sharma_requirements_2014}.
\end{itemize}

\subsection{Automating RE Tasks With Generative AI}

\textbf{LLMs} have significantly improved how we automate software engineering tasks and what quality we can achieve. Besides the improved performance over traditional ML algorithms, LLMs are mainly successful in SE because they do not need extensive datasets to be trained.
Generative LLMs build upon a decoder-only architecture derived from the well-known transformer architecture~\cite{Vaswani17}. Decoder-only LLMs have been designed to generate text. To support their generative capabilities, they are primarily pre-trained with a next-word prediction (NWP) objective, where the models predict the next word or words in a given sequence of words. Once a model has been trained, it generates text based on a so-called \textit{prompt}. A prompt is a textual input instructing the generative LLM to generate the desired response.

LLMs have been employed in various SE tasks. Existing surveys acknowledge that the application of LLMs in RE is particularly scarce compared to other software engineering areas, such as testing, code generation, and program repair~\cite{Hou2024,Fan2023}. In RE, LLMs have been investigated, e.g., to summarize legal texts~\cite{Jain2023}, in requirements traceability~\cite{rodriguez2023prompts,Vogelsang2025-smells}, for requirements classification~\cite{Zadenoori2025}, or for model generation~\cite{ferrari2024}.
A comprehensive guideline on how to apply LLMs for RE tasks can be found in a recent textbook on NLP for RE~\cite{Vogelsang2025-LLM}.

The primary input for LLMs consists of so-called prompts, which instruct the model to perform a specific task and supply the context or data required to complete it. To improve the quality of LLM outputs, various strategies have been developed to refine how prompts are constructed and formulated. Within the field of prompt engineering, numerous techniques for optimizing prompts have been proposed and their effectiveness remains an active area of research. In this study, we focus on two such techniques: zero-shot prompting, and least-to-most prompting.

\textbf{Zero-shot prompting} involves instructing an LLM to perform a specific task without providing any explicit examples, referred to as \textit{shots}, demonstrating how to accomplish it. This approach leverages the extensive pre-training of LLMs on large and diverse datasets, which gives them with the contextual knowledge needed to handle a wide range of tasks without additional task-specific training.

\textbf{Least-to-most prompting} is a prompting strategy introduced by Zhou et al.~\cite{zhouLeasttoMostPromptingEnables2023} that improves LLM performance on complex problems by decomposing them into simpler subproblems. Unlike standard chain-of-thought prompting, which provides a single, linear reasoning trace, least-to-most prompting asks the model to solve easier subproblems first and then use their solutions as context for harder ones.

\section{Methodology}

Aiming to answer the question on whether LLMs can effectively automate the requirements elicitation interview process, we explored three different methods. First, we examined two different prompting strategies, zero-shot prompting, and least-to-most prompting. Finally, we investigated whether \textit{fine-tuning} a pretrained LLM is a viable strategy for the task at hand.

In the following, the zero-shot approach will be referred to as \textit{LLMREI-short}, while the least-to-most prompting approach will be referred to as \textit{LLMREI-long}, reflecting its use of a longer and more detailed prompt.

\subsection{Zero-Shot Prompting}

Initially, we tested a zero-shot approach using GPT-4o, as it represents the simplest prompting strategy and thus serves as a natural starting point for the first experiments. The objective of this method was to ascertain whether a rudimentary prompt would already be sufficient to guide the bot through a requirements interview without the necessity for extensive customization. This approach was adopted as an initial strategy, as the reduced length of zero-shot prompts and thus the smaller number of tokens required in the LLM, results in a more efficient utilization of resources. Moreover, there is no need for the creation of examples and the following decision-making process regarding which examples to include, as is the case in, for example, few-shot prompting.

Initially, a role-based prompt was tested to establish a clear interviewer identity. However, early attempts revealed that the bot was still acting like an assistant, frequently requesting user input rather than taking the lead in the interview. To address this issue, an additional instruction was added to direct the bot to actively ask business- and project-related questions. Although this increased engagement, the bot would sometimes overwhelm users by asking multiple questions at once. The final refinement instructed the bot to behave like a real-world interviewer by asking one question at a time or two related questions per topic. We ultimately defined a concise, three-sentence system prompt to serve as the foundation for the zero-shot approach.

\begin{system*}{LLMREI-short}
You are an interviewer, called LLMREI, who assists a requirements engineer in eliciting requirements. Bombard the stakeholder with questions about his/her business and his/her project to find out everything the stakeholder envisions! Act like a real-world interviewer, so only ask one question at a time or only ask two questions if it is about one specific topic.
\end{system*}

\begin{figure}
    \centering
\includegraphics[width=1\linewidth]{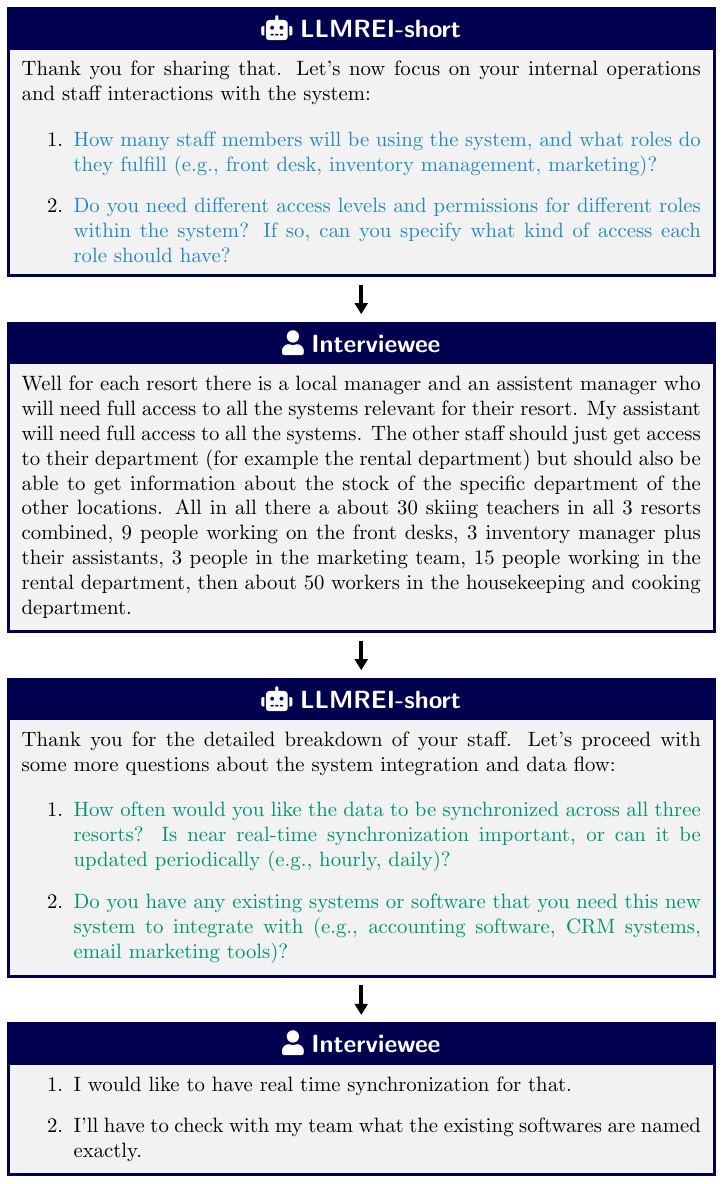}
    \caption{Conversation excerpt from an interview conducted by the LLMREI-short bot, highlighting its ability to transition smoothly between probing (blue) and follow-up (green) questions.}
    \label{fig:zero-shot-example}
\end{figure}

When tested with GPT-4o, this concise prompt demonstrated promising behavior. The bot conducted interviews in a manner closely resembling that of a human interviewer, proactively asking relevant questions and maintaining context. Figure~\ref{fig:zero-shot-example} presents an example interaction using the zero-shot bot, \textit{LLMREI-short}, illustrating its ability to transition smoothly between probing and follow-up questions. This structured dialogue balances open-ended and specific inquiries while preserving clarity and focus. This observation indicates that even a minimal yet carefully designed prompt enables GPT-4o to perform effectively as an interview bot without requiring extensive customization.

\subsection{Least-To-Most Prompting}
The second approach we investigated involved iterative prompt refinement, with the objective of assessing whether the promising results of the zero-shot approach could be improved by gradually enhancing the prompt with more explicit guidance on conducting elicitation interviews.

We began with a minimally guided prompt to test the model’s baseline capabilities. However, early trials revealed that the bot struggled to maintain focus and conduct effective interviews, indicating a need for a more structured and supportive prompt design. Rather than applying least-to-most prompting in the strict sense, where assistance increases dynamically within a single task, we adopted a session-level refinement process. With each iteration, we incrementally introduced clearer instructions and domain-specific cues based on observed weaknesses.

Our first prompt version showed potential but exhibited several issues that required refinement. We followed an iterative development cycle: after testing each version in stakeholder interviews, we analyzed transcripts and categorized mistakes using Bano~et~al.'s framework~\cite{bano_teaching_2019}. Revisions were then informed by Ferrari~et~al.'s requirements interview guidelines~\cite{ferrari_sapeer_2020}, which emphasize best practices such as discussing existing systems, involving multiple stakeholders, and summarizing key points.

We embedded Ferrari et al.'s principles into actionable instructions, evolving the prompt from a simple, zero-shot format into structured interview guidelines. The enhanced bot effectively introduced the interview, prioritized key topics at appropriate moments, and provided a coherent concluding summary. Our iterative refinement followed Ferrari~et~al.'s domain-specific recommendations \cite{ferrari_learning_2019}.

Early iterations showed steady improvement, but as we progressed, refinements became harder to assess due to the small sample size of interviews and variability in stakeholder responses. Evaluating the bot's performance became challenging since we had to act as both interviewees and reviewers, making error identification more subjective. Ultimately, we developed a structured prompt that integrates step-by-step interview guidelines, predefined roles, clear sequences for follow-up questions (i.~e., question pathways), and scenario-based examples.

During testing, we discovered that the bot often asked multiple questions simultaneously, overwhelming stakeholders. Since Bano~et~al.'s mistake categories~\cite{bano_teaching_2019} did not cover this issue, we explicitly instructed the bot to ask only one question at a time unless two were closely related. This adjustment was introduced to prevent the bot from overwhelming interviewees and to make the interview easier to follow. While not derived from existing guidelines for bot-based requirements interviews, this modification reflects our goal of creating a more manageable and user-friendly interaction based on observations during testing.

The final LLMREI-long system prompt was structured around three key components:
\begin{enumerate}
    \item \textbf{Role Description:} Clearly defining the bot's role as a requirements elicitation assistant.
    \item \textbf{Interview Guidelines:} A five-step structured process guiding the bot through interviews.
    \item \textbf{Error Handling and Behavior Adjustments:} Ensuring adaptability and professional conduct.
\end{enumerate}
An excerpt, excluding the comprehensive interview guidelines and error handling for the sake of brevity, is listed below. The full prompt can be found as part of our replication package.\footnotemark[1]

\begin{system*}{LLMREI-long}
You are an interviewer, called LLMREI, who assists a requirements engineer in eliciting requirements. \\

Goal: The purpose of this chat bot is to conduct comprehensive and effective requirements elicitation interviews with stakeholders, ensuring all necessary information is gathered to support project development. \\

\{Interview Cookbook\} \\

Maintain professionalism throughout the interview.
Adjust questions based on the stakeholder's role, education level, and domain knowledge.
Adapt questioning style to fit the flow of the stakeholder’s responses.
Actively listen to differentiate between stated needs and actual needs.
Let the customer create scenarios.
Example: ''Please visualize the first page of your application and explain how you would interact with it step-by-step.''
\end{system*}

\subsection{Fine-Tuning}
In addition to using the aforementioned prompt engineering techniques, we also experimented with fine-tuning an LLM using a dataset of requirements interview transcripts annotated with labels identifying interviewer mistakes. We used the data from Ferrari~et~al.'s~\cite{ferrari_learning_2019} study who trained university students to conduct requirements interviews with tutors acting as fictional stakeholders. A subset of students received mistake-based training, which improved their performance in later interviews. However, since the interviewers were students rather than professionals, the quality of the dataset for training purposes must be considered carefully.

Since OpenAI only supported fine-tuning on GPT-3.5 Turbo at the time of our study, we used this model, which may partly explain its poorer performance compared to the prompt engineering approaches. Preparing the interview recordings for fine-tuning also required extensive preprocessing.

\textbf{Data Preprocessing:}
We converted Ferrari et al.'s 70 interview recordings into a format usable for our purposes. After filtering out recordings lacking in quality, we retained 50 interviews. We standardized all files to the MP3 format for consistency.

We transcribed the interviews using WhisperX~\cite{Bain2023}, which integrates OpenAI's Whisper model~\cite{radford23a} with PyAnnote.audio for speaker diarization. This step helped distinguish different speakers. However, transcription accuracy varied. Some interviews had precise transcriptions, while others required manual correction for up to half the dialogue due to overlapping speech or unclear audio. Transcriptions retained speech artifacts like stutters and repetitions, preserving authenticity but potentially introducing noise into fine-tuning.

To align with OpenAI's best practices~\cite{openai_dataset}, we used our refined long prompt as the system prompt during fine-tuning. The final training dataset consisted of the long prompt and the transcribed interviews.

\textbf{Results and Conclusion:}
We fine-tuned three versions of GPT-3.5 Turbo:
\begin{compactitem}
    \item \textbf{Model A:} Fine-tuned with all 50 interview transcripts.
    \item \textbf{Model B:} Fine-tuned with an 80/20 train-validation split of the 50 interviews.
    \item \textbf{Model C:} Fine-tuned using only the 10 best interviews, as identified in Ferrari~et~al.'s evaluation~\cite{ferrari_sapeer_2020}.
\end{compactitem}

Testing these models with the long prompt revealed disappointing results. Models A and B, which used the full dataset, performed worse than Model C, which was trained on the highest-quality interviews. Even more concerning, all fine-tuned models underperformed compared to LLMREI-long and LLMREI-short, which used the long and short prompts without fine-tuning.

Several problems emerged from evaluating the results: the fine-tuned bot produced incoherent responses, struggled to ask clear questions, and sometimes lost track of the interview's purpose. In one instance, it unexpectedly switched to a different language. Given these flaws, we decided against further evaluation with participants, as the non-fine-tuned models clearly outperformed the fine-tuned ones.

The primary cause of poor performance appears to be the mixed quality of  data used for fine-tuning. The student interviewers lacked professional experience, with many of them being non-native English speakers. As a result, their interviews lacked the depth and structure of professional requirements interviews. Additionally, training the model on spoken-language transcripts likely caused the resulting bot to mimic informal and unstructured dialogues, further degrading performance.

Our results highlight the challenges associated with fine-tuning LLMs using domain-specific data of limited quality. Given the suboptimal performance and the already promising results achieved with the prompt engineering approaches, we decided not to pursue the fine-tuning strategy further and excluded it from our evaluation.

\section{Evaluation}
The primary objective of the evaluation was to assess the performance of the two LLM-based requirements interview bots, LLMREI-long and LLMREI-short, in conducting requirements elicitation interviews. To address this objective, a series of interviews were conducted using both bots in an experimental setup.

\subsection{Research Questions}

\begin{compactitem}
\item \textbf{RQ1:} To what extent can LLMREI minimize common errors that typically occur in traditional requirements interviews?
\item \textbf{RQ2:} How effective is LLMREI in eliciting relevant information from interviewees?
\item \textbf{RQ3:} To what extent can LLMREI adapt its questions based on the content and context provided by the interviewee?
\end{compactitem}

\subsection{Experiment Setup and Data Collection}

We designed an experiment involving real interviews, where participants acted as stakeholders in predefined scenarios. Each participant assumed the role of a stakeholder in one of two scenarios previously used in the study by Ferrari~et~al.~\cite{ferrari_sapeer_2020}: (1) a hair and nail salon seeking a digital solution for managing appointments and employee scheduling, or (2) a ski resort requiring a digital booking and business management platform for its three locations. These scenarios formed the foundation of the interviews and were kept consistent to enable a comparative analysis between the two chat bot configurations. The full scenarios are available in our replication package\footnotemark[1] and will be referred to as the \textit{Ski} and \textit{Salon} scenarios throughout the paper.

The interview process followed a structured approach. Before each session, participants received a description outlining their role as a stakeholder and instructions on what to do after the interview. The LLMREI chat bot then conducted the interview. Upon completion, participants were asked to fill out a questionnaire evaluating the quality of the interview and the effectiveness of the questions posed by the chat bot.

To facilitate the interviews, we developed a website that implemented this structured process. Participants accessed the website, where they were randomly assigned a scenario and provided with its description. They then engaged in the interview with the chat bot through a chat-based interface. After the interview, each participant received a PDF containing another interview for review. The interviews were conducted in person by students during two university courses. Table~\ref{tab:study-cases} provides an overview of the number of completed interviews per scenario and prompt.


\begin{table}
    \centering
    \caption{Number of interviews per scenario and prompt}
    \label{tab:study-cases}
    \begin{tabular}{@{}lrrr@{}}
    \toprule
         \textbf{Scenario}& \textbf{Long prompt} & \textbf{Short prompt} & \textbf{Sum}\\
         \midrule
         Salon & 6 & 9 &\textbf{15} \\
         Ski & 11 & 7 &\textbf{18}\\
         \midrule
         \textbf{Sum} & \textbf{17} & \textbf{16} \\
         \bottomrule
    \end{tabular}
\end{table}

We used GPT-4o (precisely, \textit{gpt-4o-2024-08-06}) using the OpenAI API. Depending on which bot was used, either the short or long prompt was set as the system prompt, with all other settings left unchanged. Each interview lasted around 30 minutes, including the time taken by the participants to read the scenario description, and to complete the questionnaire. The transcripts of all interviews were stored in CSV format and were later converted into PDF documents to provide
participants with a readable version for completing their reviews.

The participants were primarily computer science students between the ages of 20 and 30. All participants were native German speakers, though proficient in English, which was the language intended to be used for the interviews. Some participants switched to German for their interviews, which was handled flawlessly by the LLM, showing its adaptability to different target groups.

For the second research question, which focuses on the accuracy of the elicited requirements, a ground truth of requirements based on the scenarios was created by us. We reviewed each interview manually (two reviewers) to obtain a list of elicited or partially elicited requirements and compared the results to reach a consensus.

\subsection{Data Analysis and Metrics}

\textbf{RQ1 (Quality of Interviews):}
To address RQ1, we focused on common mistake types identified in requirements interviews to create our questionnaire. These mistake types are derived from the study of Bano~et~al.\cite{bano_teaching_2019} who categorized common mistakes made during requirements interviews. Some of the common mistakes are not appropriate for assessing LLMREI because they are about human characteristics, such as the ''voice of analyst not [being] clear'', or they are about external factors, such as teamwork and planning. Therefore, these types of mistakes are not
considered. The mistakes suitable to assess the LLMREI and their categories are shown in Figures \ref{fig:rq1-bar-chart-long-vs-human}, and \ref{fig:rq1-bar-chart-short-vs-human}.

Building on this, Ferrari et al.~\cite{ferrari_learning_2019} developed a questionnaire based on these mistakes to evaluate interviewer performance with one question corresponding to exactly one of the mistakes. We adapted this questionnaire for our study, omitting certain questions not applicable to an interview bot environment. Each of the displayed mistake types corresponded to one question in the questionnaire. The mistake groups
considered for our evaluation are:
\begin{compactitem}
    \item Question Formulation: Errors related to how questions are phrased, potentially leading to misunderstandings or incomplete information.
    \item Question Omission: Failure to ask necessary questions to gather comprehensive requirements.
    \item Order of Interview: Issues arising from the sequence in which questions are asked,
    affecting the flow and clarity of the interview.
    \item Communication Skills: Errors due to ineffective communication techniques.
    \item Customer Interaction: Mistakes related to engaging with the customer.
\end{compactitem}

Responses were collected using a Likert scale ranging from~1~(Strongly disagree) to 5 (Strongly agree), where a lower score indicates a lower occurrence of the mistake. This scoring system allows for a quantitative assessment of the presence of mistakes across different interview scenarios. This scale was also used by Ferrari~et~al. in their study.
All questions used and their corresponding mistake type and mistake group can be seen in Figures~\ref{fig:rq1-bar-chart-long-vs-human}, and~\ref{fig:rq1-bar-chart-short-vs-human}. To provide a meaningful baseline for evaluating the effectiveness of LLMREIs,
we used data from the study conducted by Ferrari~et~al.~\cite{ferrari_learning_2019}. In their study, university students, some of whom had received mistake-based training in requirements elicitation interviews, conducted interviews while being assessed using the same mistake types and questions in the questionnaire, we used in our study.
We focused specifically on 18 student-led interviews conducted after mistake-based training, as this subset provided a more appropriate comparison group, considering the interviewers had at least some formal training in conducting requirements elcitation interviews. This targeted approach ensures that our comparison is grounded in relevant and appropriately filtered data, allowing us to assess how LLMREIs perform relative to trained human interviewers.

Our study tested several settings to compare the effectiveness of LLMREIs against traditional interviews and to explore variations within LLMREIs themselves. The settings
include:
\begin{compactitem}
    \item LLMREI vs.\ Human: Comparing the performance of LLM-based interviews with that of human interviewers.
    \item Prompt Variations (LLMREI-long vs.\ LLMREI-short): Evaluating the impact of prompt complexity on the effectiveness of LLMREI.
\end{compactitem}

The main focus of our evaluation is on the comparison between the LLMREI-long and LLMREI-short. This comparison aims to determine how the depth and detail of prompts influence the reduction of common errors in the interview process.

In our evaluation, both individual mistake types and mistake groups were analyzed to provide a more granular and comprehensive assessment of LLMREIs. By first examining individual mistake types, we could identify specific areas where LLMREIs either succeeded
or struggled in mitigating common mistakes. Additionally, by grouping mistakes into broader categories, we gained higher-level insights that allowed us to assess overarching trends in LLMREI performance.

\textbf{RQ2 (Effectiveness):}
The data used to address this research question was collected by filling out questionnaires after each interview. These questionnaires specifically
aimed to answer the question: Was the necessary and relevant information elicited?
To achieve this, each requirement from the ground truth (a predetermined list of requirements based on the interview scenario) was evaluated. The responses to the questionnaire were structured into three categories:
\begin{compactitem}
    \item Not elicited (1): The requirement was not mentioned or implied during the interview.
    \item Partially elicited (2): Some aspects of the requirement were addressed, but the full detail was not obtained.
    \item Fully elicited (3): The requirement was thoroughly discussed and understood.
\end{compactitem}

We reviewed only the summary of each interview to assess these categories.

The ground truth of requirements was established by all authors of this study. By examining the interview scenarios, we derived a comprehensive list of requirements essential to each scenario. The Cool Ski Resort scenario contained 12 requirements, whereas the Hair Salon scenario contained 8.

\textbf{RQ3 (Adaptability):} 
To answer RQ3, we manually analyze all interviews and categorize all questions raised by LLMREI into one of the following categories:
\begin{enumerate}
    \item \textbf{Context-independent:} General questions that, while relevant, could be drawn from a finite set of standardized questions common in requirements interviews. This category shows \textbf{no adaptation capabilities}; a simple list of questions could be used instead.
    \item \textbf{Parametrized:} General questions that, while relevant, could be drawn from a finite set of standardized questions that are parameterized with some basic context information such as name of the project. This category shows \textbf{low adaptation} capabilities; A simple non-AI tool could also provide such questions (e.g., by entering project information into question templates with predefined placeholders)
    \item \textbf{Context-deepening:} Questions that are context-specific, directly informed by the interviewee's prior answers or the specifics of the scenario. This category shows \textbf{medium adaptation} capabilities to the context.
    \item \textbf{Context-enhancing:} Questions that elaborate on the context and add information that was not mentioned before. This category shows \textbf{high adaptation} capabilities to the context including bringing up new (creative) ideas.
\end{enumerate}

We report the distribution of categories across all interview questions. A high ratio of context-deepening and context-enhancing questions would indicate that LLMREI can adapt its questions to the interviewer's responses.  

\subsection{Threats to Validity}
\textbf{Internal Validity:}
A key threat to internal validity is selection bias, as participants were primarily students with varying experience levels in requirements engineering. More experienced participants may have provided more structured feedback and engaged in deeper discussions, influencing results. To mitigate this, we derived real-world scenarios from Ferrari~et~al.'s \cite{ferrari_learning_2019} work to ensure relevance to software engineering projects. Participants were also briefed on interview expectations to simulate stakeholder behavior more realistically. A threat that still remains is the lack of \textit{real} interest in the provided system.

\textbf{External Validity:}
The study focused on software engineering scenarios, limiting generalizability to other industries. Future research should assess the LLMREI bot in different domains. Another limitation was participant demographics---most were students or young professionals, which does not reflect the diversity of real-world stakeholders. Including more varied experience levels and backgrounds in future studies would provide deeper insights into the bot's effectiveness.
Additionally, the study used simplified interview scenarios, which covered generic and popular types of applications. While sufficient for initial testing, these do not capture the complexities of real-world requirements elicitation, where systems may be domain-specific or highly innovative. Expanding the bot's application to more complex settings will be necessary to evaluate its broader applicability.

\textbf{Construct Validity:}
The study assessed interview quality through participant questionnaires, which involved subjective evaluations of mistakes. While we used a standardized evaluation framework to ensure consistency, subjective bias remained a concern. To minimize variability, we structured feedback in an objective format, reducing personal biases in evaluation.
Another issue was applying human-centric mistake categories to evaluate the bot. These categories were designed for human interviewers and may not fully capture LLM-specific errors. Future research should refine mistake classifications to better align with AI behavior.
The study also relied on students acting as stakeholders, which raises concerns about representation. Without deep domain knowledge, their responses may differ from those of real professionals. A future study with actual stakeholders would help validate findings.
Finally, our customization methods focused on LLMREI-long and LLMREI-short prompts, potentially overlooking other effective prompt engineering techniques. Future research should explore alternative approaches.

\textbf{Conclusion Validity:}
With only 33 interviews conducted, the small sample size limits statistical significance. A larger dataset would enhance result reliability and reduce the likelihood of random variation affecting conclusions.

\subsection{Study Results}

\subsubsection{RQ1 (Quality of Interviews)}

\begin{figure}
    \centering
    \includegraphics[width=\linewidth]{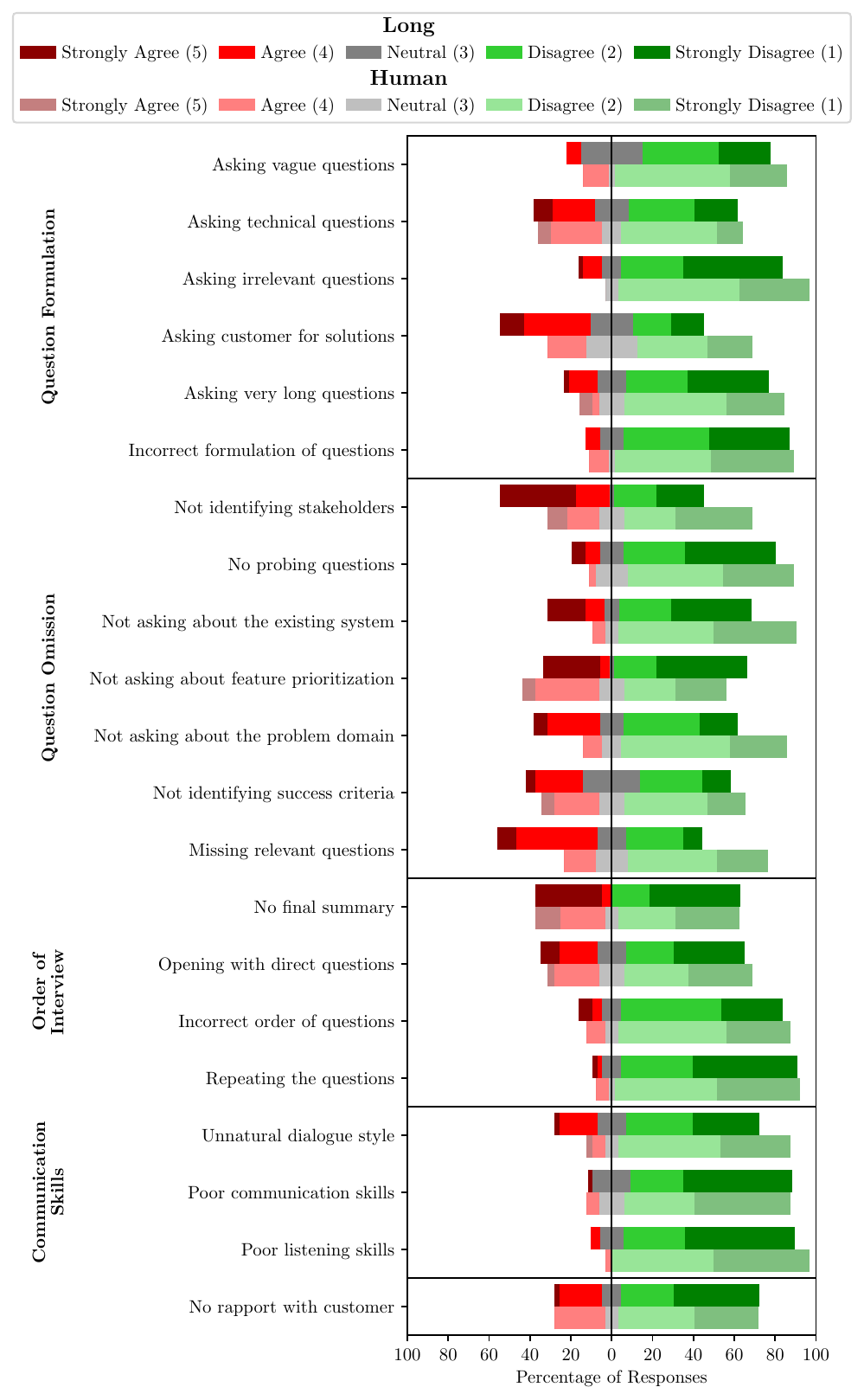}
    \caption{Evaluation of common mistakes made by the \textbf{LLMREI-long} bot, and human interviewers. For each mistake, the upper bar shows the rating for LLMREI, while the lower bar shows the rating for the human interviewers. A score of 5 (agree) means that the rater found the mistake to be highly prevalent, while 1 (disagree) means that the mistake was not recognized in the interview.}
    \label{fig:rq1-bar-chart-long-vs-human}
\end{figure}

\begin{figure}
    \centering
    \includegraphics[width=\linewidth]{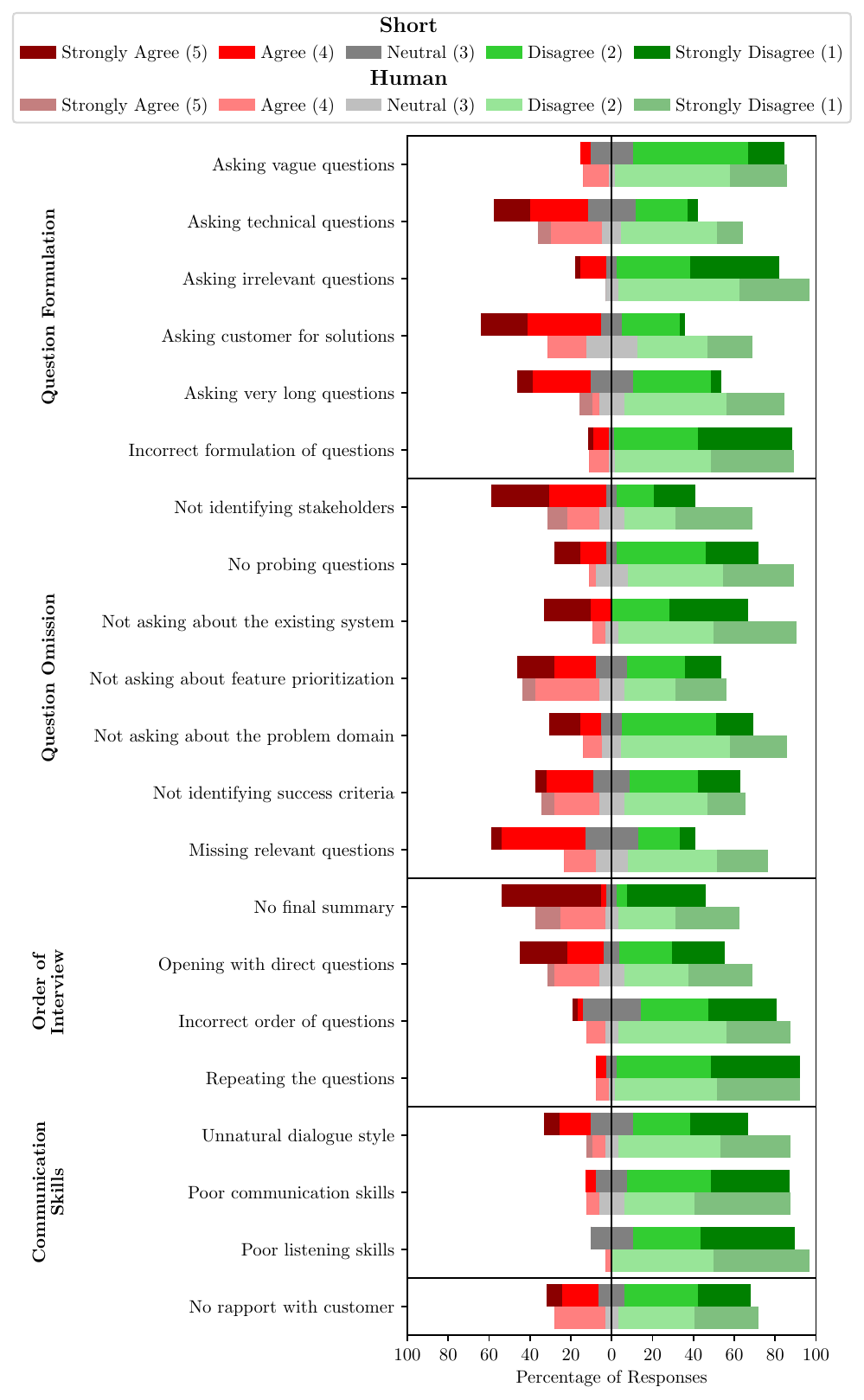}
    \caption{Evaluation of common mistakes made by the \textbf{LLMREI-short} bot, and human interviewers. For each mistake, the upper bar shows the rating for LLMREI, while the lower bar shows the rating for the human interviewers. A score of 5 (agree) means that the rater found the mistake to be highly prevalent, while 1 (disagree) means that the mistake was not recognized in the interview.}
    \label{fig:rq1-bar-chart-short-vs-human}
\end{figure}

We started by comparing the overall results achieved with LLMREI to the results acquired from the interviews that were completed by human interviewers. The goal was to compare the common mistakes made by the LLMREI bots with those made by the human interviewers in the study by Ferrari~et~al.~\cite{ferrari_learning_2019}. Fig.~\ref{fig:rq1-bar-chart-long-vs-human}, and Figure \ref{fig:rq1-bar-chart-short-vs-human} show the results achieved with either the long (LLMREI-long), or the short (LLMREI-short) prompt compared to those obtained from the human interviews.

As can be seen, LLMREI focused on a very methodical approach, laying, for example, a stronger focus on feature prioritization or generating final summaries than the human interviewers. Also, LLMREI showed an excellent performance in terms of communication skills, highlighting the good language understanding expected from current LLMs. Despite the minor difference, raters completing our questionnaire even estimated the communication skills to be better than those of the human interviewers.

Comparing both figures, one can see that the long prompt performed better in general than the short prompt. 64.23\% of the reviewers either strongly disagree (1) or disagree (2) with recognizing the mistakes in the interviews by the long prompt, while 59.1\% answered similarly for the short prompt. This becomes especially clear for mistakes on which guidance was included in the long prompt but missing in the short prompt. For example, the long prompt showed consistently better results in generating a final summary after completion of the interview. This can be derived from the fact that the long prompt included a remark asking the LLM to generate a summary. Also, the long prompt included information on how to ask questions, resulting in reviewers rating the long prompt to perform better in avoiding very long questions. This behavior shows that the LLM is capable of precisely following the instructions provided in the prompt, highlighting the potential for further improvement of the results by applying more in-depth prompt tuning techniques.

\subsubsection{RQ2 (Effectiveness)}

\begin{figure}
    \centering
    \includegraphics[width=\linewidth]{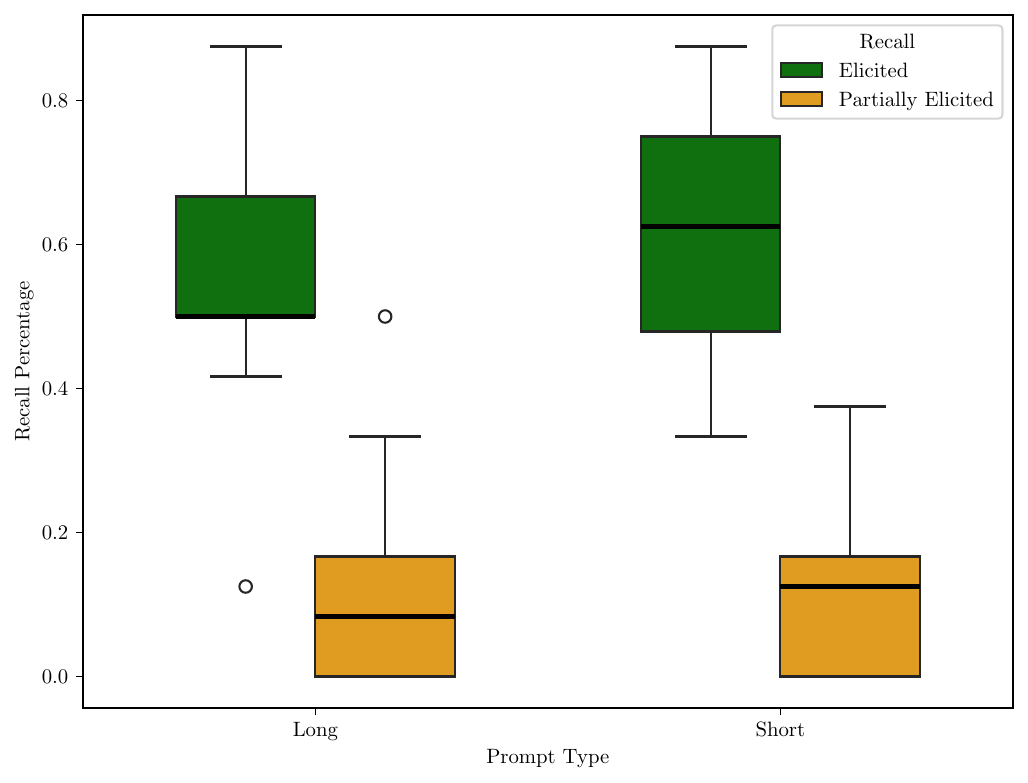}
    \caption{Ratio of fully and partially elicited requirements (i.e., \textit{recall}) using the long and short prompt.}
    \label{fig:rq2-boxplot-recall}
\end{figure}

Evaluating how many of the requirements included in the given scenarios were actually elicited in the interviews yielded that LLMREI was able to completely elicit up to 60.94~\% of all requirements and partially elicit up to 12.76~\% (in total 73.7~\%) using the short prompt which performed better at this task. Figure~\ref{fig:rq2-boxplot-recall} shows the distribution of elicited requirements over all interviews.

\begin{figure}
    \centering
    \includegraphics[width=\linewidth]{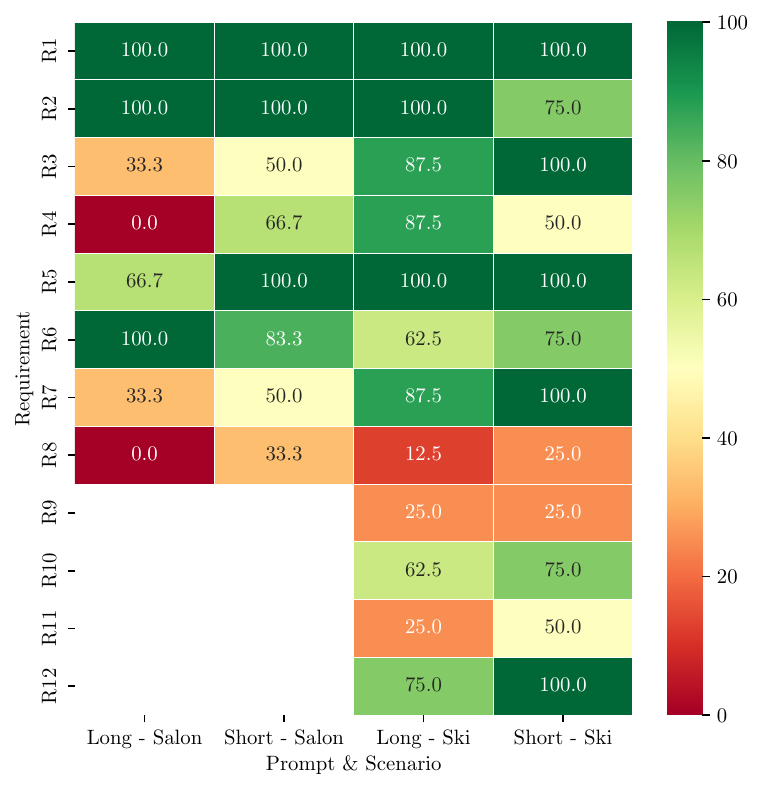}
    \caption{Percentage of interviews in which a specific requirement was elicited.}
    \label{fig:rq2-heatmap-requirements}
\end{figure}

Figure~\ref{fig:rq2-heatmap-requirements} shows in how many interviews each requirements of both scenarios was elicited using the long, and the short prompt. As can be seen, the short prompt not only performed better in eliciting a large set of requirements in general but also manages to elicit all requirements at least once over all interviews. Meanwhile, the long prompt fails to do so for two requirements of the Salon scenario. Additionally, the figure shows that requirements R1, R2, and R5 were elicited best over both scenarios (thus, being six different requirements), with all of these requirements being functional requirements.

\subsubsection{RQ3 (Adaptability)}

\begin{table}
    \centering
    \caption{Distribution of question categories in interviews}
    \label{tab:rq3-results}
    \begin{tabular}{@{}lrr@{}}
    \toprule
     \textbf{Category}    & \textbf{Short} & \textbf{Long} \\
    \midrule
     Context-independent    & 27.3\% & 28.3\%\\
     Parameterized    &  12.8\%& 28.9\%\\
     Context-deepening   & 44.4\% & 32.3\%\\
     Context-enhancing    & 15.3\% & 10.4\% \\
    \bottomrule
    \end{tabular}
\end{table}

Table~\ref{tab:rq3-results} presents the distribution of question categories across all interviews conducted using both the long and short prompts. The results indicate that, regardless of the prompt length, approximately half of the LLM-generated questions strongly depend on previous user responses.

The short prompt produced a higher number of context-enhancing questions, where the chat bot not only followed up on responses but also introduced new ideas (e.g., \textit{In terms of customer interaction, would you like any additional features, such as appointment reminders, loyalty programs, or feedback collection?}''). Similarly, it also generated more context-deepening questions, prompting users to elaborate on specific aspects of a particular feature (e.g., \textit{For managing supplies, do you want the software to track inventory levels, alert you when stock is low, or even manage orders from suppliers? How do you manage your supply level presently?}’’). This high occurrence of both context-deepening and context-enhancing questions made the interviews feel highly adaptive and responsive to user input. However, this also meant that the chat bot's suggestions occasionally influenced the direction of the conversation.

In contrast, interviews using the long prompt were more structured, with fewer context-specific elements in the questions (e.g., ``\textit{Let’s start by gathering detailed requirements for the customer portal. Could you please elaborate on what you want this online customer portal to accomplish?}’’). We classified such questions as `parameterized' since the only context-specific component is the placeholder \textit{customer portal}, while the rest of the question remains general. Although the higher proportion of parameterized questions might suggest that the long prompt has less ability to adapt to context, it could also be seen as more suitable for requirements elicitation, as it reduces the interviewer's influence on responses.

\begin{figure}
    \centering
\includegraphics[width=1\linewidth]{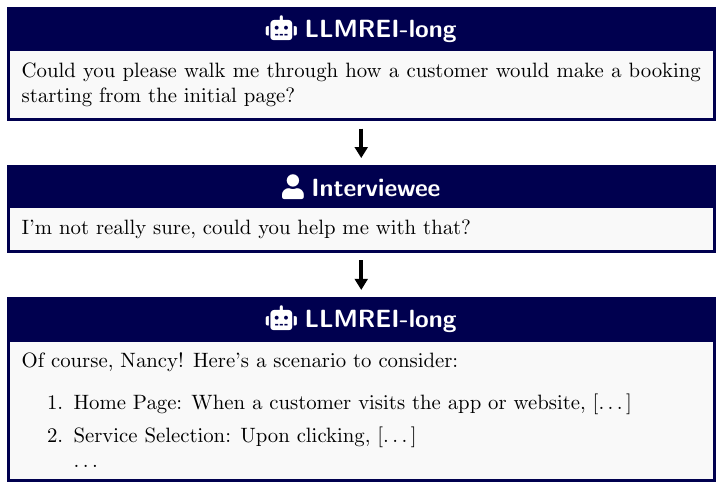}
    \caption{Conversation excerpt from an interview conducted by the LLMREI-long bot, showing a scenario where the user specifically requested guidance from the chat bot.}
    \label{fig:context-enhancing-example}
\end{figure}

In some cases, users appreciated and even requested the chat bot's context-enhancing capabilities. Several interviews featured dialogues similar to what can be seen in Figure~\ref{fig:context-enhancing-example}.

Both prompts resulted in a comparable proportion of context-independent questions. These were typically used at the beginning of the interview to establish the discussion framework (e.g., \textit{How would you describe the project we're discussing today in two to four sentences?}'') or to ask predefined wrap-up questions toward the end (e.g., ``\textit{To wrap up, could you suggest who else might have valuable insights into this project and whose needs we should consider?}''). However, we also observed that both prompt versions were highly responsive to users expressing a lack of time or interest in the interview. Rather than insisting on predefined mandatory elements---such as concluding with a summary or inquiring about additional stakeholders---the chat bot adapted by swiftly ending the conversation. While this flexibility can enhance the user experience by respecting participants' constraints, it may also result in missed opportunities to gather crucial information that would typically be covered in a structured interview process.

\section{Discussion}

The results of this study demonstrate that LLMs, particularly when configured with carefully designed prompts, can effectively support the requirements elicitation process. Our results show that LLMs can automate requirements elicitation interviews while maintaining an accuracy and quality close to that of human-led interviews. While the findings indicate that LLMs perform promisingly in this area, full automation remains a challenge. This can be concluded in terms of our research questions as follows:
\begin{compactitem}
    \item RQ1: The LLMREI-long prompt demonstrated a reduction in common interview errors, suggesting that with structured prompting, LLMs can help reduce mistakes typically made by human interviewers. Both LLMREIs showed comparable performance to human-led requirements interviews.
    \item RQ2: There was no significant difference in the number of elicited requirements between the long and short prompts. However, the LLMs facilitated the requirements elicitation process, though they may not yet consistently elicit all requirements.
    \item RQ3: Both LLMs, regardless of whether they used the long or short prompt, showed a reasonable mix of context-specific and general requirements interview questions. This suggests that the LLMs possess excellent language skills and guidance capabilities, enabling them to steer the conversation and adapt questions based on the flow of the interview.
\end{compactitem}

While LLMs may not yet fully automate the requirements elicitation process, they can serve as powerful support tools for requirements engineers, particularly in scenarios with limited human resources or large numbers of stakeholders. LLMs offer the potential for consistency and scalability in interviews, but human intervention is still needed when dealing with more complex or domain-specific requirements.
The performance of LLMREI was compared with human-led interviews and the results from Ferrari~et~al.~\cite{ferrari_learning_2019}. Although LLMs can reduce common errors, they sometimes fall short of capturing all requirements. Combining LLMREI with Surana~et~al.'s~\cite{rajender_kumar_surana_intelligent_2019} requirements elicitation automation architecture could offer further improvements, and this combination would be an interesting direction for future research. Additionally, further refinement of LLM capabilities and more sophisticated prompt designs could enhance LLMREI's overall utility in requirements elicitation.

\subsection{Limitations of Automated Interviewing}

Despite the promising results, our approach also has limitations that must be considered when interpreting the findings. The use of pre-trained LLMs inherently comes with constraints related to model behavior. One of the issues is the tendency to generate hallucinations, which could result in incorrect information being presented during interviews or in the summaries of elicited requirements. For instance, in one interview, the chat bot provided an estimated project price when asked by the user, despite having no reliable basis for such a calculation. This risk of misinformation is particularly concerning in requirements elicitation, where accuracy is essential. 

Another concern is the chat bot’s tendency to cross predefined boundaries. In one interview, it asked the user for their email address to facilitate scheduling a follow-up meeting. While this might seem like an advanced feature, it raises ethical and privacy concerns regarding the collection of personal information. Further fine-tuning the prompt may be necessary to ensure the chat bot adheres to strict data-handling guidelines and avoids overstepping privacy boundaries.

Data privacy is another significant limitation, particularly when using models like GPT. The handling of interview data remains unclear, raising concerns about the model's suitability for privacy-sensitive domains. These concerns could limit the practical application of such models in real-world settings where data protection is a priority. However, these risks may be mitigated by employing corporate or self-hosted LLMs, which offer enhanced data privacy measures and prevent user data from being stored. While this solution provides greater security, it also introduces a potential barrier to adoption, as the computational demands and infrastructure requirements could be prohibitive for some users or organizations.

Overall, these limitations underscore the need for further research to refine this approach. Future work should explore increasing the sample size, involving multiple independent reviewers, experimenting with a broader range of prompts and customization techniques, and conducting interviews with a more diverse and experienced participant pool. Addressing these challenges will be crucial in enhancing the reliability, accuracy, and applicability of LLM-driven requirements elicitation.

A fundamental limitation of LLM-based requirements interviews is their inability to capture and interpret the non-verbal cues that are essential in human-to-human communication. In traditional face-to-face interviews, body language, facial expressions, tone of voice, and subtle shifts in hesitation or emphasis provide valuable contextual information that helps interviewers gauge confidence, detect uncertainty, and probe deeper into critical areas. These cues can reveal implicit needs, conflicting priorities, or emotions that respondents may not explicitly verbalize. In contrast, text-based interactions with an LLM rely solely on written or spoken words, stripping away these rich layers of communication. As a result, important nuances may be lost, leading to less effective clarification, misinterpretations, or missed opportunities to explore underlying concerns. This limitation makes LLM-based interviews less suitable for highly sensitive or complex elicitation scenarios where interpersonal dynamics play a crucial role in uncovering unspoken requirements.

\section{Conclusions}
This paper explored the potential of LLMs in automating requirements elicitation interviews. The study evaluated three approaches: zero-shot prompting, least-to-most prompting, and fine-tuning. The findings showed that LLMREI, especially when using the least-to-most prompting method, performed similarly to human-led interviews in many respects, particularly in minimizing common errors. However, fully automating the requirements elicitation process remains a challenge, as the LLMs in our experiment only captured a bit more than 70\% of the intended requirements.
Despite these limitations, LLMREI shows promise as a valuable tool for requirements engineers, potentially reducing their workload and improving consistency across interviews. Future work should focus on scaling these approaches to real-world projects, finding more refined prompts, and exploring new methods to enhance LLM adaptability.

\bibliographystyle{IEEEtran}
\bibliography{references}

\end{document}